\documentclass[a4paper,11pt]{article}
\usepackage{epsfig,subfigure}
\usepackage{cite}
\date{}
\begin{document}
\title{
\Large{
\textbf{Alignment transition in a nematic liquid crystal due to
field-induced breaking of anchoring}
}
}
\author{
Valentina S. U. Fazio\footnote{e-mail: \texttt{fazio@fy.chalmers.se}},
and Lachezar Komitov\\
\textit{\small{Liquid Crystal Physics, Department of
Microelectronics \& Nanoscience}}, \\
\textit{\small{Chalmers University of Technology
\& G\"oteborg University,}}\\
\textit{\small{S-41296 G\"oteborg, Sweden}}
}
\maketitle
\begin{abstract}
\noindent
We report on the alignment transition of a nematic liquid crystal from 
initially homeotropic to quasi-planar due to field-induced anchoring 
breaking. 
The initial homeotropic alignment is achieved by Langmuir-Blodgett
monolayers.
In this geometry the anchoring strength can be evaluated by the 
Frederiks transition technique.
Applying an electric field above a certain threshold provokes
turbulent states denoted DSM1 and DSM2.
While DSM1 does not affect the anchoring,
DSM2 breaks the coupling between the surface and the liquid 
crystal: switching off the field from a DSM2 state does not immediately
restore the homeotropic alignment.
Instead, we obtain a quasi-planar metastable alignment.
The cell thickness dependence for the transition is related to the 
cell thickness dependence of the anchoring strength.
\end{abstract}

\noindent
Pacs {61} {30$-$v} {Liquid crystals}\\
Pacs {78} {20.Jq} {Electrooptical effects}\\
Pacs {47} {27Cn} {Transition to turbulence}\\

Nematic liquid crystals (NLCs) can be aligned along a well defined
direction either by treating the confining surface in a certain 
way\cite{Jerome91}, or by applying an external electric (or magnetic) 
field\cite{deGennes}.  
In the absence of an external field the director $\mathbf{n}$
coincides with the so-called \lq\lq easy axis\rq\rq \,whose orientation
depends on the surface treatment.  
An additional electric (or magnetic) field can,
depending on the LC's dielectric (magnetic) anisotropy, destabilize this 
orientation and the director $\mathbf{n}$ will deviate from the easy 
axis (Frederiks transition).  
Depending on the strength of that field, the anchoring is 
said to be weak or strong.

At higher fields, above certain thresholds, modulated structures
due to convective instabilities appear
which are characterized by a periodic distortion of the nematic director
$\mathbf{n}$ along certain directions and which can 
be visualized in form of regular patterns.
At still higher electric fields  
one finds a second and a third threshold that
correspond to the appearance of two different turbulent states called
DSM1 and DSM2, where DSM stands for \lq\lq dynamic scattering mode\rq\rq 
\, because the director $\mathbf{n}$ undergoes a chaotic motion which leads
to intense scattering of light.

These instabilities have been extensively studied for planar and
homeotropically oriented MBBA, which is a NLC with negative 
dielectric anisotropy.  
We have found particularly interesting and helpful for our present
study the work of Versace \textit{et al.} 
\cite{ScaVerCar95,CarLucScaVer95,CarScaVer97} 
who have investigated the transition between DSM1 and DSM2
states and suggested that this transition may be
related to a breaking of the surface anchoring.  
We note, however, that their experiments have been performed in
planar MBBA cells of fixed thickness.

In this work we present a detailed experimental study on the effects 
of the turbulent states on the homeotropic nematic anchoring and present
direct evidence of field-induced breaking of the anchoring.
The use of homeotropically aligned MBBA cells has the  
important advantage that in this geometry one can obtain the independent 
information about the surface anchoring strength from the Frederiks 
transition threshold. 
This is not possible in planar MBBA cells as the electric 
field stabilizes the alignment in this case.

Recently we have shown \cite{FazKomLag98a,FazKomLag98b} 
that a Langmuir-Blodgett (LB) monolayer of stearic acid (C18) can be used 
to obtain a very good and stable homeotropic alignment of MBBA.  
LB monolayers of C18 have been deposited on ITO coated glass plates 
which have then been used to make sandwich cells of 4, 8, 10, and 15\,$\mu$m
thickness.
The samples were observed in a polarising microscope which was connected 
to an image acquisition system.
A sinusoidal electric field was applied across the cells,
perpendicular to the glass plates, and the optical response was 
monitored on the oscilloscope via a photodetector. 
We defined as threshold field the one at which 10\,\% of the
saturated value of the transmitted light intensity was reached.
All investigations have been performed at room temperature.

At relatively weak electric fields the Frederiks transition takes place.  
Since MBBA has negative dielectric anisotropy
($\Delta \varepsilon = -0.7$\cite{deGennes}) the molecules want to 
orient perpendicular to the electric field, which is in conflict
with the initial homeotropic alignment.
The threshold field was found to be frequency independent\cite{EL1note2}.

From the electric field threshold it is possible to calculate\cite{Sonin} 
the anchoring strength $w$ of the NLC on the surface according to the equation:
\begin{displaymath}
w = q K_{3} \tan \left( \frac{q d}{2} \right),
\,\,\,\,\,\,\,\,\,\,
\mbox{with}
\,\,\,\,\,\,\,\,\,\,
q = E \sqrt{\frac{|\Delta \varepsilon| \varepsilon_{0}}{K_{3}}},
\end{displaymath}
where $d$ is the cell thickness, $\Delta \varepsilon$ is the
dielectric anisotropy, $\varepsilon_{0}$ is the vacuum dielectric
constant, $K_{3}$ is the bend elastic constant, and $E$ is the 
electric field threshold at which the Frederiks transition
takes place.

\begin{figure}
\begin{center}
\epsfig{file=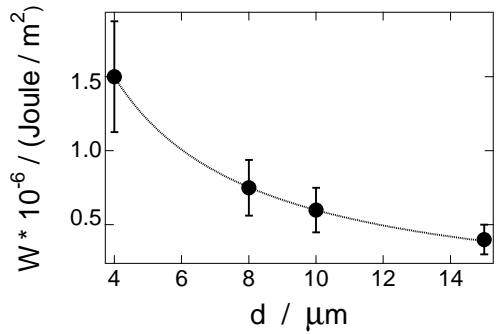,width=0.5\textwidth} 
\end{center}
\caption{
\label{figuno}
\small{
Dots: anchoring strength of MBBA on LB monolayers of C18 as a function of
the cell thickness. Dashed line: fit of the experimental values to the
function in eqn. \protect\ref{eqnuno}. 
The parameters of the best fit are: $w_{R} = 1.25 \times 
10^{-5}$\,Joule/m$^{2}$ and $\Sigma = 1.3 \times 10^{-4}$\,C/m$^{2}$.}
}
\end{figure}

The anchoring strength was found to depend on the cell thickness.
The values are shown in Figure \ref{figuno}.
%
According to Barbero \textit{et al.}\cite{AleBarPet93,BarDur90} 
the thickness dependence of the effective anchoring strength is 
related to the presence of the conductive charges in the NLC volume
and to the adsorption of charges at the boundaries. 
The actual distribution of charges in the external electric field 
depends on the cell thickness and due to the flexoelectric properties 
of NLCs the effective anchoring energy also appears to be thickness 
dependent. 
The effective anchoring strength can be written as:
\begin{equation}
w_{eff} = w_{R} + \frac{\sigma}{\varepsilon \varepsilon_{0}} \left[
\frac{\Delta \varepsilon}{2 \varepsilon} \lambda_{D} \sigma + 2 e
\right],
\label{eqnuno}
\end{equation}
where $w_{R}$ is the  anchoring strength in the Rapini-Papoular\cite{Sonin} 
model, $\varepsilon$ is the average dielectric constant
($\varepsilon$ = 5.05\cite{deGennes}), 
$\lambda_{D}$ is the Debye screening length
($\lambda_{D}$ = 200\,nm\cite{BarPet94}), $e$ is the 
average flexoelectric coefficient ($e = - 1.3 \times 10^{-12}$\,C 
m$^{-1}$
\cite{BlinovChigrinov}), and
\begin{displaymath}
\sigma = \Sigma \frac{d}{d + 2 \lambda_{D}}
\end{displaymath}
is the absorbed charge density at the boundary, where
$\Sigma$ depends on the conductivity of the LC and on the
characteristics of the surfaces.  
We have fitted the data in Figure \ref{figuno} to the eqn. (\ref{eqnuno}) 
and found that $w_{R}$ is positive as it should be for an initially 
homeotropically oriented NLC\cite{Sonin}.

\begin{figure}
\begin{center}
\epsfig{file=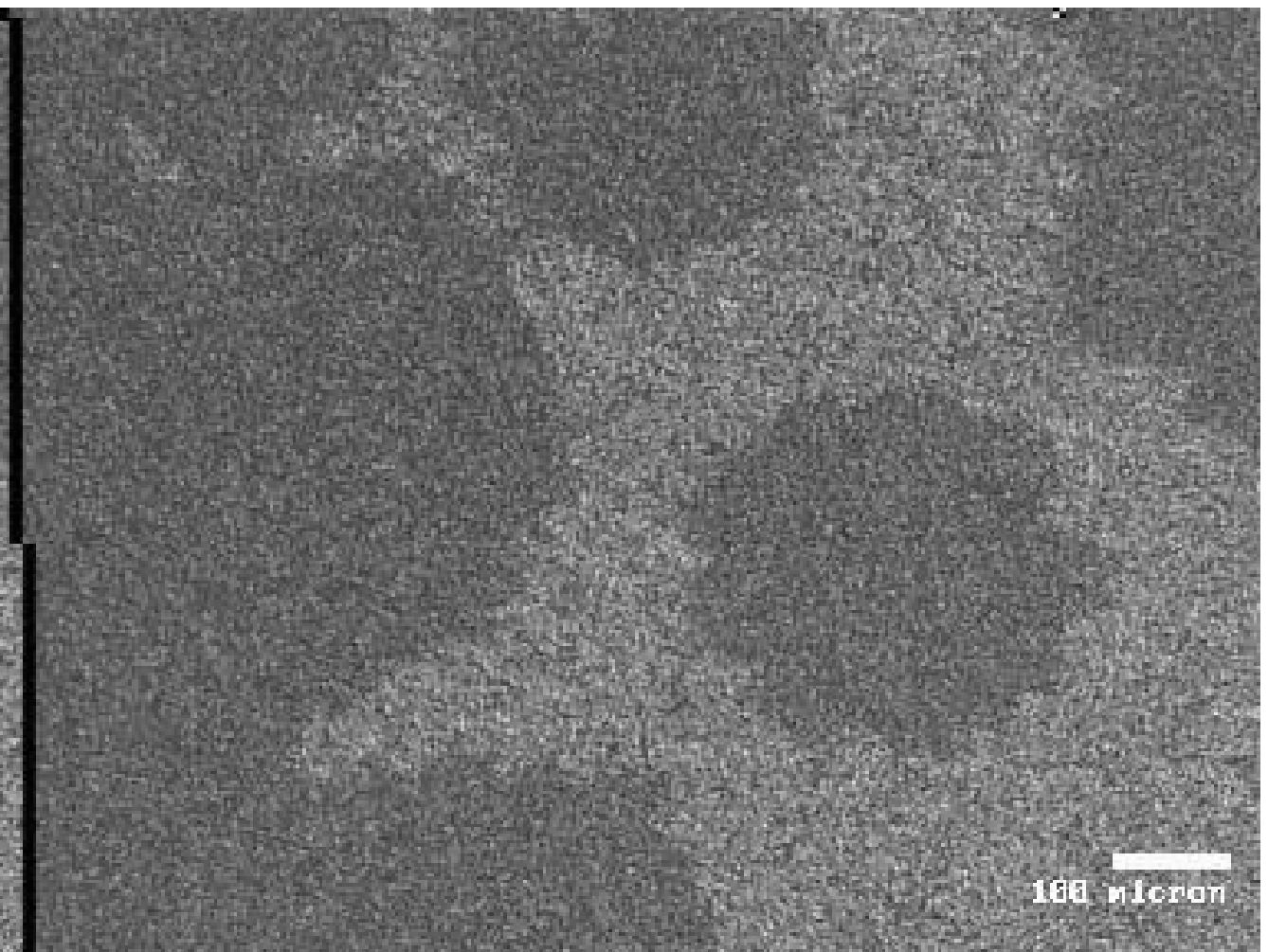,width=0.4\textwidth}
\hspace{0.05\textwidth}
\epsfig{file=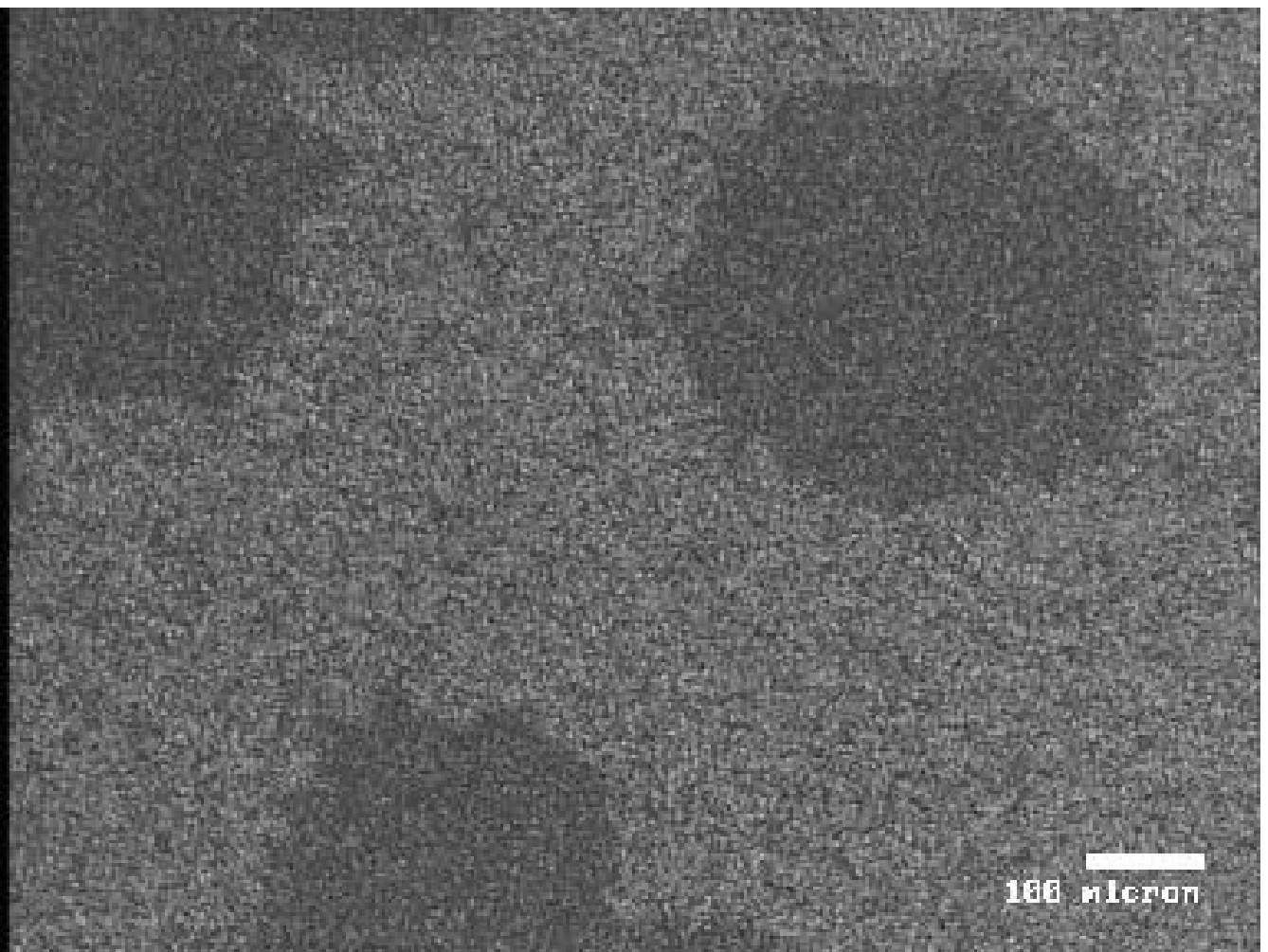, width=0.4\textwidth}
\end{center}
\caption{
\label{figcinque}
\small{DSM2 domains expand in the DSM1 turbulent state.
Note the circular form of the DSM2 domains which is due to the
absence of a preferred direction in the plane of the cell given by
the initial homeotropic alignment.  
} 
}
\end{figure}

Just above the Frederiks transition threshold the NLC is essentially
quasi-planar oriented.
On increasing the field above the Frederiks transition threshold 
we observed first modulated structures (Williams rolls),
then their destabilization, and then 
the transition to the weakly turbulent state DSM1
(for more details about electrohydrodynamic instabilities in 
NLCs see for instance \cite{BlinovChigrinov}, chapter\,5). 
Switching off the electric field from these states the homeotropic 
alignment is immediately restored.
Increasing the electric field further above the DSM1 threshold,
we observed the transition to the second turbulent state, DSM2.
Domains of DSM2 nucleate at some points in the nematic layer, 
which is in the DSM1 state, and 
expand to the whole sample (see Figure \ref{figcinque}).

We would like to point out that the electric field was increased very
slowly, so that the Frederiks transition as well as the modulate 
structures and the turbulences could fully develop.

DSM1 and DSM2 have different optical appearance: in both states the 
light is scattered very strongly, but in DSM2 the size of the 
scattering centers is much smaller than in DSM1, and DSM2 
domains are darker than DSM1 domains between crossed polarizers.
%
We note that in the case of planar MBBA, the DSM2 domains expand
more quickly in the aligning direction\cite{ScaVerCar95}.  
In our case they expand in all directions with the same speed because 
the initial homeotropic alignment does not create a preferred direction in
the plane of the cell.
Hence, the domains are circular.  

\begin{figure}
\begin{center}
\epsfig{file=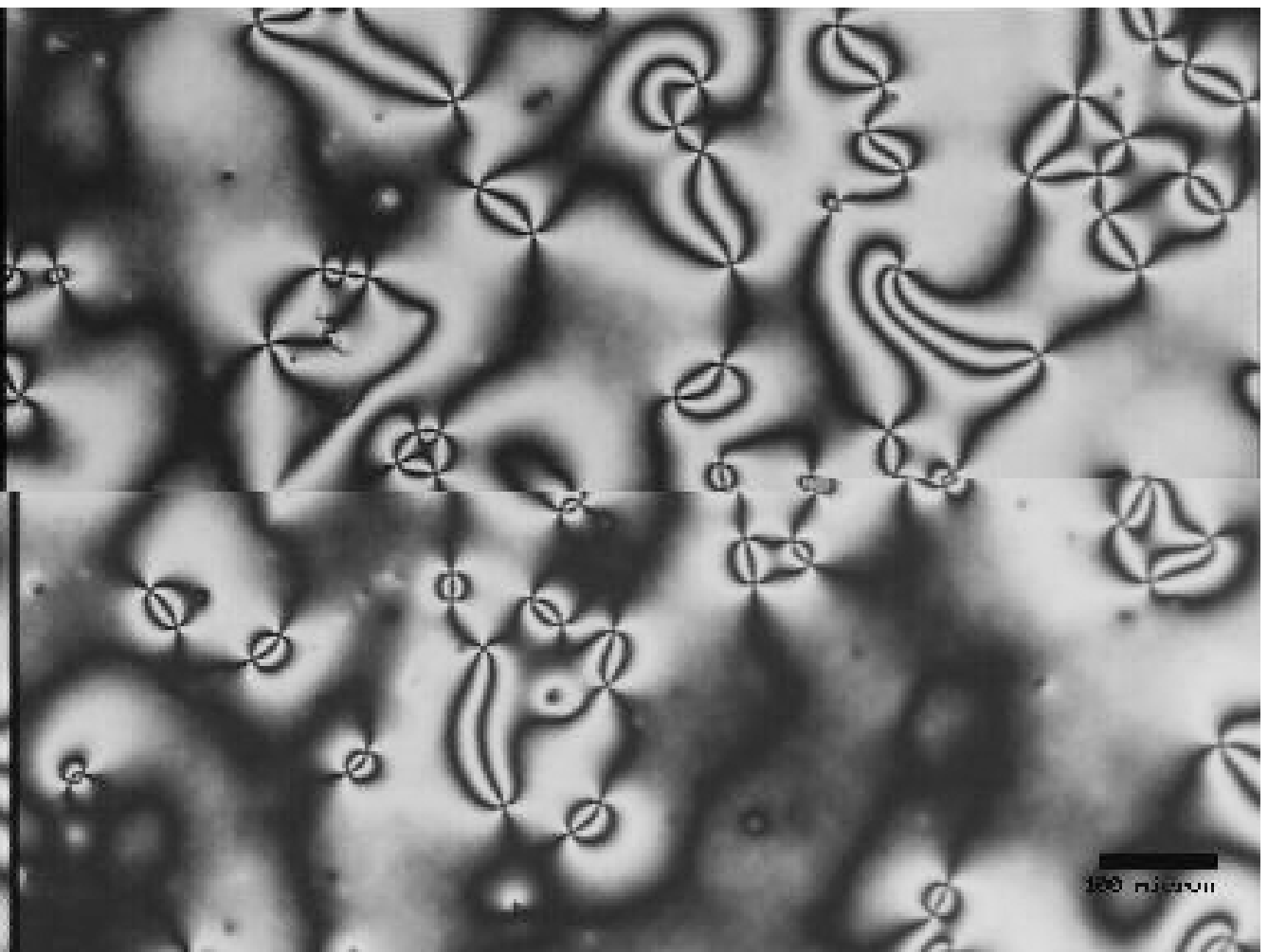, width=0.3\textwidth}
\hspace{0.01\textwidth}
\epsfig{file=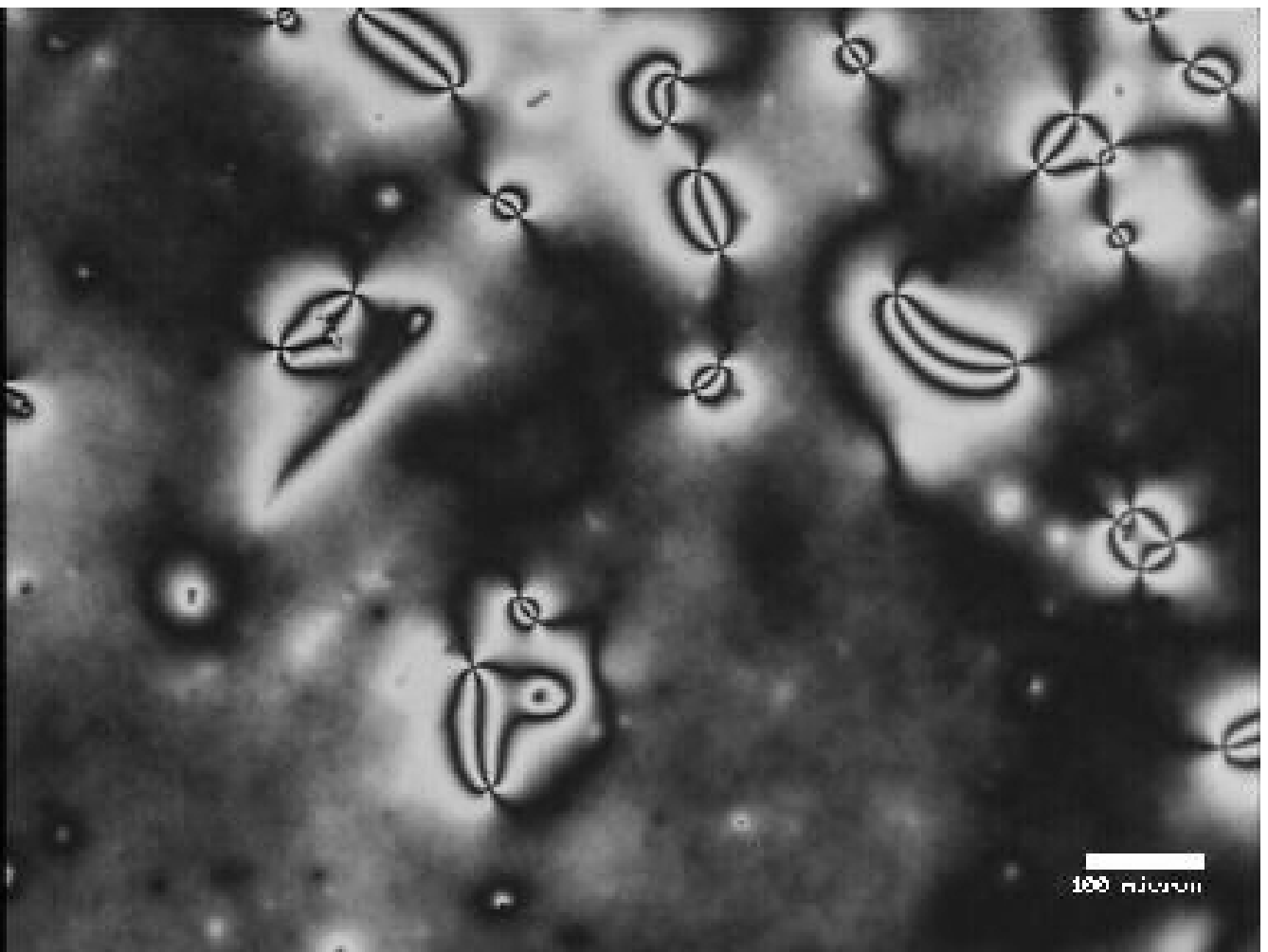, width=0.3\textwidth}
\hspace{0.01\textwidth}
\epsfig{file=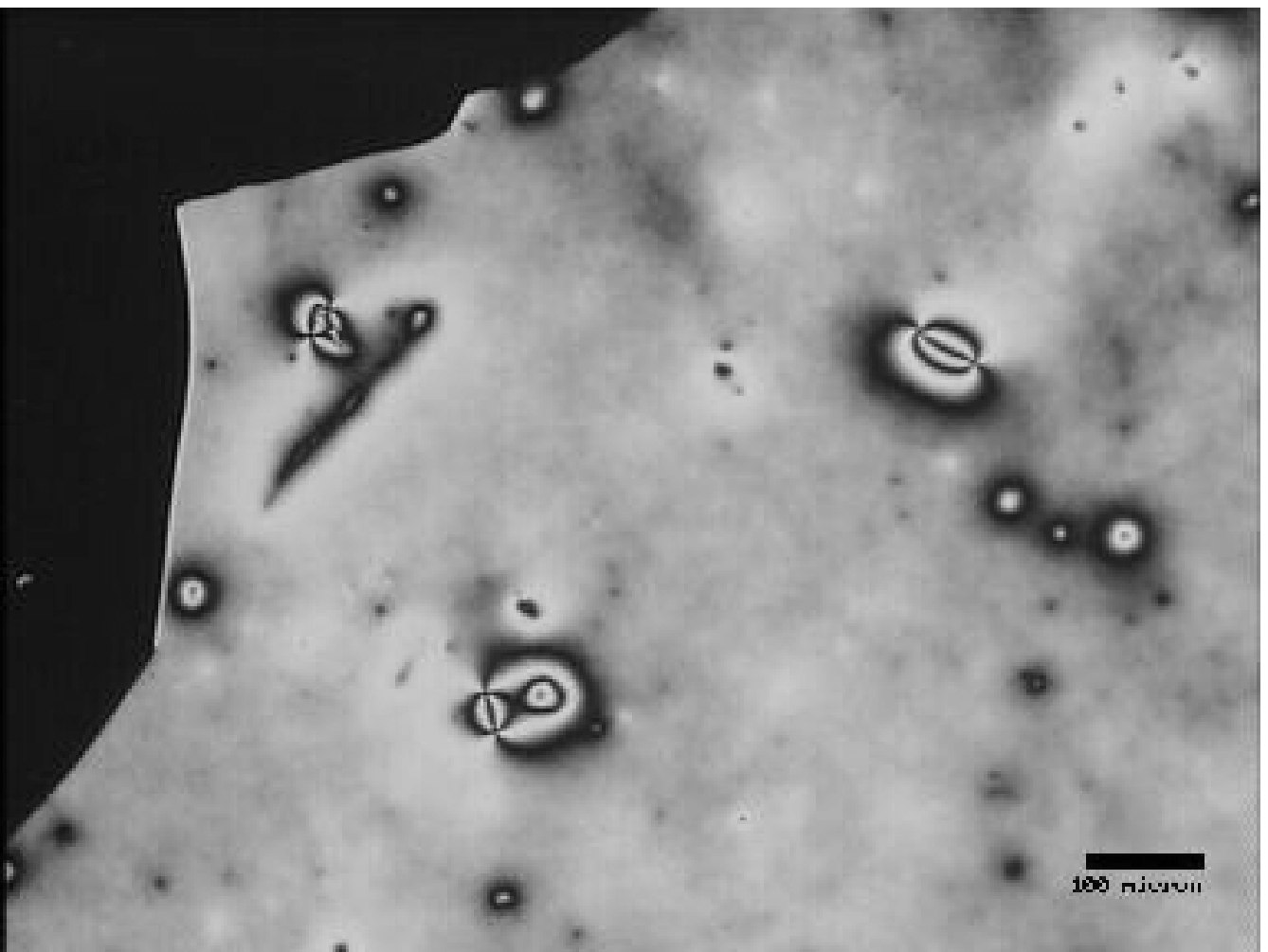, width=0.3\textwidth}
\end{center}
\caption{
\label{figcinqueA}
\small{Evolution of alignment.
Left: t=0; the electric field has been switched off and the sample 
does not return to the homeotropic alignment; instead,
a metastable state appears which presents many defects.
Center: t=10\,s; the defects disappear and leave the sample in a
quasi-planar state.
Right: t=30\,s; the defects have almost disappeared and the NLC 
adopts a planar alignment with the director oriented
along the filling direction; the dark domain is a
homeotropic domain which expands into the quasi-planar one until the whole
sample becomes homeotropic again.
Cell thickness 4\,$\mu$m.
} 
}
\end{figure}

Switching off the electric field from the DSM2 state does
not restore immediately the homeotropic alignment.  
Instead, the sample contains many defects (see Figure \ref{figcinqueA}) 
that disappear rather quickly and leave the NLC in a metastable 
quasi-planar state. 
%
In other words, DSM2 causes a breaking of the anchoring of the 
NLC to the surface resulting in an alignment transition from 
homeotropic to quasi-planar.

\begin{figure}
\begin{center}
\epsfig{file=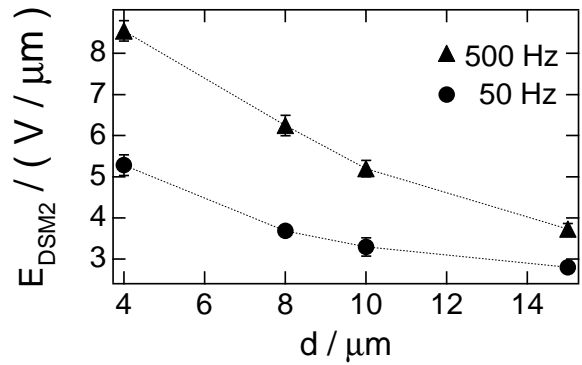, width=0.6\textwidth} \end{center} 
\caption{
\label{figsei}
\small{Electric field thresholds for the appearance of
the DSM2 turbulent state as a function the cell thickness in the 
conductive regime ($f = 50\,$Hz) and in the dielectric regime
($f = 500\,$Hz).
}
}
\end{figure}

The electric field thresholds for the onset of the DSM2 state are shown in
Figure \ref{figsei} for both the conductive and the dielectric 
regime \cite{BlinovChigrinov, EL1note1}.
%
From Figure \ref{figsei}
one can readily see that the threshold field for DSM2
depends on the cell thickness, such that the threshold is higher 
in thinner cells.
This cell thickness dependence of $E_{DSM2}$ 
can be directly related to the cell thickness 
dependence of the anchoring strength.
That is, if the anchoring strength decreases with $d$, a lower 
threshold will be needed to break the anchoring in thicker cells,
which is indeed what we observe. 

In the quasi-planar state in which the sample is left after switching 
off the electric field the NLC molecules 
are oriented in the direction of the NLC flow during the cell filling,
while the LB deposition direction does not play any role.  
As mentioned above, the quasi-planar state is metastable:
domains of homeotropic orientation nucleate with time at the
edges of the cell or in the proximity of impurities and expand
until the whole sample becomes homeotropic again\cite{FazKomLag98a}.

\vspace{4mm}

\textit{Conclusions} --
We have studied
an electric field induced alignment transition due to the 
breaking of the surface anchoring in initially homeotropic NLCs.
In this context it was important to determine the
anchoring strength within the same set of experiments as the 
electric field thresholds for the turbulent state DSM2 which is responsible 
for the breaking of the anchoring. 
The initial homeotropic alignment made it possible to study the two
phenomena independently, which could not have been done in
\cite{ScaVerCar95,CarLucScaVer95,CarScaVer97} because of the initial
planar alignment of the NLC in that case.

The DSM1-to-DSM2 transition is affected by the surface anchoring.
Versace \textit{et al.}\cite{ScaVerCar95} have assumed that the DSM1-to-DSM2
transition is governed by a surface tilting transition induced by the
external electric field: below the DSM1 threshold the nematic director 
deformation is confined to a plane; at the DSM1 transition point the 
director goes out of the plane, but this three-dimensional deformation 
reaches the surface only above the threshold of the DSM2 state, leading to 
anchoring breaking.
This explanation is in agreement with our experimental results which 
show that the system does not immediately return to the initial 
homeotropic state after switching off the field in the DSM2 state.
On the other hand, the DSM1 turbulent state does not affect the anchoring. 
Moreover, the cell thickness dependence of the electric field threshold
for DSM2 follows the behavior of the cell thickness dependence of the
anchoring strength, which is again a direct evidence of anchoring breaking. 

\bibliography{journal2,EL1.bib}

\begin{thebibliography}{10}
\bibitem{EL1note2}
The ions present in the liquid crystal bulk can influence the reorientation
  process at the Frederiks transition treshold. This effect, however, is very
  small (about a few percents) and could not be resolved by our equipment (note
  that the magnitude of the error bar in Figure \protect\ref{figuno} is about
  15-20\,\%).
\bibitem{EL1note1}
Electrohydrodynamic instabilities (EHIs) are very sensitive to the frequency of
  the applied field\cite{BlinovChigrinov}. We can distinguish two different
  threshold behaviors. For frequencies $f$ less than a certain critical
  frequency $f_{c}$ the threshold electric field $E_{W}$ for the appearance of
  the EHIs increases sharply with $f$ (conductive regime). For frequencies $f >
  f_{c}$ (dielectric regime) $E_{W} \propto \sqrt{f}$. We have measured the
  threshold electric fields for the appearance of the EHIs as a function of the
  frequency and estimated the critical frequency to be $f_{c} \approx $
  100\,Hz.
\end{thebibliography}
\end{document}